\documentclass[conference]{IEEEtran}
\IEEEoverridecommandlockouts
\usepackage{cite}
\usepackage{amsmath,amssymb,amsfonts}
\usepackage{algorithmic}
\usepackage{graphicx}
\usepackage{textcomp}
\usepackage{xcolor}
\usepackage{hyperref}
\def\BibTeX{{\rm B\kern-.05em{\sc i\kern-.025em b}\kern-.08em
    T\kern-.1667em\lower.7ex\hbox{E}\kern-.125emX}}

\usepackage{caption}

\usepackage{fancyhdr}
\begin{document}

\title{Assessing FAIRness of the\\Digital Shadow Reference Model\\
\thanks{Funded by the Deutsche Forschungsgemeinschaft (DFG, German Research Foundation) under Germany's Excellence Strategy – EXC-2023 Internet of Production – 390621612.}
}

\author{\IEEEauthorblockN{Johannes Theissen-Lipp}
\IEEEauthorblockA{\textit{Chair of Databases and Information Systems} \\
\textit{RWTH Aachen University}\\
Aachen, Germany \\
theissen-lipp@dbis.rwth-aachen.de\\
ORCID: \href{https://orcid.org/0000-0002-2639-1949}{0000-0002-2639-1949}}
}

\maketitle

\begin{abstract}
Models play a critical role in managing the vast amounts of data and increasing complexity found in the IoT, IIoT, and IoP domains. The Digital Shadow Reference Model, which serves as a foundational metadata schema for linking data and metadata in these environments, is an example of such a model. 
Ensuring FAIRness (adherence to the FAIR Principles) is critical because it improves data findability, accessibility, interoperability, and reusability, facilitating efficient data management and integration across systems. 

This paper presents an evaluation of the FAIRness of the Digital Shadow Reference Model using a structured evaluation framework based on the FAIR Data Principles. Using the concept of FAIR Implementation Profiles (FIPs), supplemented by a mini-questionnaire, we systematically evaluate the model's adherence to these principles. Our analysis identifies key strengths, including the model's metadata schema that supports rich descriptions and authentication techniques, and highlights areas for improvement, such as the need for globally unique identifiers and consequent support for different Web standards. The results provide actionable insights for improving the FAIRness of the model and promoting better data management and reuse. This research contributes to the field by providing a detailed assessment of the Digital Shadow Reference Model and recommending next steps to improve its FAIRness and usability.
\end{abstract}

\begin{IEEEkeywords}
FAIR Data Principles, FAIRness, Internet of Production, Reference Model, Digital Shadow
\end{IEEEkeywords}

\section{Introduction}

In today's interconnected world, effective data management is critical, especially in the areas of Internet of Things (IoT), Industrial Internet of Things (IIoT), and emerging Internet of Production (IoP)~\cite{Pennekamp.2019.TowardsanInfrastructureEnablingtheInternetofProduction,brauner2022computer}. These fields generate massive amounts of data that must be efficiently captured, stored, processed, and shared to realize their full potential~\cite{brockhoff2021process}. Without robust data management strategies, organizations face challenges such as data silos, inefficient operations, and limited interoperability, all of which can hinder innovation and scalability. As a result, ensuring that data systems are designed to support effective communication and (re)usability across multiple parties is essential to succeed~\cite{Theissen-Lipp:SemanticsInDataspaces,behery2023actionable}.

One way to address data management challenges is through the use of reference models~\cite{Bader.2020.TheInternationalDataSpacesInformationModelAnOntology}. Reference models serve as standardized frameworks that provide structure and a common understanding among various stakeholders, including developers, data providers, data consumers, and decision makers. These models facilitate communication by providing clear guidelines and common terminology, reducing misunderstandings, and promoting collaboration. The Digital Shadow Reference Model~\cite{Becker.2021.AConceptualModelforDigitalShadowsinIndustry,Michael.2023.ADigitalShadowReferenceModelforWorldwideProduction} is an example of such a framework, focusing on the representation and use of Digital Twins or Shadows of physical assets in a structured and consistent manner. By defining key concepts, interactions, and data flows, it supports the efficient and systematic use of digital assets.

However, ensuring the proper (re)use of data and services within such frameworks imposes additional requirements. The FAIR Data Principles\cite{Wilkinson2016} -- Findability, Accessibility, Interoperability and Reusability -- promise to address these issues. These principles aim to improve the usability of data by others, whether human or machine, and provide a foundation for efficient data sharing and integration. By aligning with these principles, data systems become more discoverable, accessible, and compatible, enabling greater value to be extracted from the data. Achieving high \emph{FAIRness} is critical to ensuring that the Digital Shadow Reference Model not only serves its immediate purpose, but also facilitates the broader goal of making digital artifacts more reusable by diverse stakeholders.

This paper aims to assess the FAIRness of the Digital Shadow Reference Model by systematically evaluating its alignment with the FAIR Data Principles. Through this analysis, the paper identifies key benefits of the model in enabling effective data management and highlights existing gaps where improvements are needed. Finally, the paper provides actionable recommendations for evolving the Digital Shadow Reference Model to achieve greater FAIRness. These improvements aim to make the model and its associated artifacts more reusable, thereby promoting greater interoperability and utility in IoT, IIoT, and IoP systems.

\section{Background}

\subsection{The Digital Shadow Reference Model}

The \emph{Digital Shadow Reference Model}~\cite{Becker.2021.AConceptualModelforDigitalShadowsinIndustry,Michael.2023.ADigitalShadowReferenceModelforWorldwideProduction} was developed within the Cluster of Excellence \emph{Internet of Production}\footnote{\url{https://iop.rwth-aachen.de}} and provides a structured approach to representing Digital Twins or Shadows of physical assets. While these two concepts (twin vs. shadow) are often used interchangeably, a key difference is that Digital Twins typically include a feedback loop to the real world (i.e., actuators or control mechanisms), while Digital Shadows do not, but instead focus more deeply on deriving insights.
Digital Shadows can serve as digital counterparts that reflect the states, behaviors, and attributes of real-world entities, enabling real-time monitoring, decision making, and analysis. By providing a standardized framework, the Digital Shadow Reference Model helps ensure consistency in the way data about physical assets is collected, stored, and shared among stakeholders. This model is particularly important in the IoT, IIoT, and IoP context~\cite{Pennekamp.2019.TowardsanInfrastructureEnablingtheInternetofProduction}, where seamless communication and interoperability across heterogeneous systems and participants is critical.

The model defines key components and interactions that are helpful for the creation and utilization of digital shadows. These include data sources (e.g., sensors), data integration pipelines, and interfaces for data consumers (e.g., analytics systems or other Digital Twins/Shadows). By aligning these elements, the Digital Shadow Reference Model provides a foundation for interoperable digital ecosystems.

\subsection{The FAIR Principles}

The FAIR Data Principles~\cite{Wilkinson2016} - Findability, Accessibility, Interoperability, and Reusability - were introduced to improve the management and use of digital assets\footnote{See \url{https://www.go-fair.org/fair-principles/} for a handy overview.}. These principles aim to ensure that data is:

\begin{enumerate}
    \item \textbf{Findable}: Easily discoverable through standardized identifiers and rich metadata.
    \item \textbf{Accessible}: Available through well-defined access protocols that maintain data integrity and security.
    \item \textbf{Interoperable}: Compatible with other data and systems via shared standards and formats.
    \item \textbf{Reusable}: Well-described and licensed to enable broad reuse across contexts.
\end{enumerate}

Meeting these principles enhances the usability of data for both humans and machines, and promotes seamless integration and collaboration. The FAIR Principles are widely recognized as best practices in data management~\cite{sds1,sds2} and are increasingly being adopted across industries and dataspaces\cite{franklin2005databases,otto2019international,sds3}.

In parallel with initiatives around \emph{FAIR Digital Objects}~\cite{FDO1} as driver for efficient data exchange\footnote{Notable drivers are the \href{https://fairdo.org/}{FDO Forum} and the \href{https://www.bundesdruckerei.de/de/innovation-hub/was-sind-fair-digital-objects}{German Federal Printing Office}.}, there exist efforts to accelerate broad community convergence on FAIR implementation options.
One of these efforts are \emph{Reusable FAIR Implementation Profiles} (FIP)~\cite{Schultes2020}, which include a mini-questionnaire for creating an own FAIR implementation profile.
In this paper, we utilize this mini-questionnaire for evaluating the FAIRness of the Digital Shadow Reference Model by adding an evaluation column ranging from \texttt{-{}-} to \texttt{++} (see Table \ref{table:findings}.

\subsection{Why Align Reference Models with FAIR Principles?}

Integrating the FAIR Principles into reference models, such as the Digital Shadow Reference Model, is beneficial to addressing key challenges in modern data ecosystems. Thus alignment ensures that the artifacts managed within the model - such as data, metadata, and services - become more FAIR and thus are more easily discoverable, accessible, and reusable. This is especially important in typical IoT / IIoT / IoP domains where large volumes of data are generated and shared among multiple stakeholders.

Compliance with the FAIR Principles promotes interoperability, enabling digital assets to function seamlessly across different systems and organizations. It also ensures that these assets can be effectively reused in different contexts, maximizing their value. For the Digital Shadow Reference Model, adopting the FAIR Principles not only improves its practical utility, but also enhances its scalability and relevance in a data-driven worldwide production lab.

\section{Methodology}

Evaluating the FAIRness of the Digital Shadow Reference Model requires a structured assessment framework based on the four pillars of the FAIR Data Principles: Findability, Accessibility, Interoperability, and Reusability. To this end, the concept of \emph{FAIR Implementation Profiles} (FIPs)~\cite{Schultes2020} provides a robust foundation for this evaluation. The FIP framework, supplemented by a mini-questionnaire\footnote{\url{https://bit.ly/yourFIP}} and a best practice guide\footnote{\url{https://www.go-fair.org/how-to-go-fair/fair-implementation-profile/}}, systematically breaks down each FAIR principle into specific criteria.  

This mini-questionnaire consists of 21 focused questions designed to measure compliance with the FAIR Principles. Using this framework, we systematically assess the FAIRness of the Digital Shadow Reference Model, identifying its strengths and weaknesses through detailed responses to each criterion. Table~\ref{table:fair_criteria} shows the complete list of FIP mini-questionnaire questions that serve as our evaluation criteria.

\begin{table*}[tbhp]
\caption[Assessment criteria for FAIR principles from the FIP mini-questionnaire derived from~\cite{Schultes2020}.]{Assessment criteria for FAIR principles from the FIP mini-questionnaire\protect\footnotemark derived from~\cite{Schultes2020}.}
\begin{center}
\begin{tabular}{|l|l|l|}
\hline
FAIR Principle & No. & Question                                                                                                    \\ \hline
F1             & \#1 & What globally unique, persistent, resolvable identifiers do you use for metadata records?                   \\ \hline
F1             & \#2 &What globally unique, persistent, resolvable identifiers do you use for datasets?                           \\ \hline
F2             &\#3 & Which metadata schemas do you use for findability?                                                          \\ \hline
F3             &\#4 & What is the technology that links the persistent identifiers of your data to the metadata description?      \\ \hline
F4             &\#5 & In which search engines are your metadata records indexed?                                                  \\ \hline
F4             &\#6 & In which search engines are your datasets indexed?                                                          \\ \hline
A1.1           &\#7 & Which standardized communication protocol do you use for metadata records?                                  \\ \hline
A1.1           &\#8 & Which standardized communication protocol do you use for datasets?                                          \\ \hline
A1.2           &\#9 & Which authentication \& authorisation technique do you use for metadata records?                            \\ \hline
A1.2           &\#10 & Which authentication \& authorisation technique do you use for datasets?                                    \\ \hline
A2             &\#11 & Which metadata longevity plan do you use?                                                                   \\ \hline
I1             &\#12 & Which knowledge representation languages (allowing machine interoperation) do you use for metadata records? \\ \hline
I1             &\#13 & Which knowledge representation languages (allowing machine interoperation) do you use for datasets?         \\ \hline
I2             &\#14 & Which structured vocabularies do you use to annotate your metadata records?                                 \\ \hline
I2             &\#15 & Which structured vocabularies do you use to encode your datasets?                                           \\ \hline
I3             &\#16 & Which models, schema(s) do you use for your metadata records?                                               \\ \hline
I3             &\#17 & Which models, schema(s) do you use for your datasets?                                                       \\ \hline
R1.1           &\#18 & Which usage license do you use for your metadata records?                                                   \\ \hline
R1.1           &\#19 & Which usage license do you use for your datasets?                                                           \\ \hline
R1.2           &\#20 & Which metadata schemas do you use for describing the provenance of your metadata records?                   \\ \hline
R1.2           &\#21 & Which metadata schemas do you use for describing the provenance of your datasets?                           \\ \hline
\end{tabular}
\label{table:fair_criteria}
\end{center}
\end{table*}

Data for the assessment is gathered from descriptions of the model, its usage within the \emph{Internet of Production} project (use case interviews), documentation, general references on the Digital Shadow Reference Model, and literature on the FAIR Principles and related best practices. Note that some of these resources are still only available internally in the said project and have not yet been published for the scientific public. These sources provide the necessary context and insight to assess the alignment of the model with the FAIR Principles.

The evaluation procedure is conducted in three stages:
\begin{itemize}
    \item \textbf{Mapping the Model to FAIR Principles}: The components, processes, and outputs of the Digital Shadow Reference Model are systematically mapped to the criteria outlined in the assessment framework.
    \item \textbf{Scoring}: Each criterion is scored based on qualitative or quantitative metrics, such as the presence or absence of specific features and the degree of alignment with the FAIR Principles.
    \item \textbf{Analysis}: The results are synthesized to identify patterns, including areas of strength and opportunities for improvement.
\end{itemize}

While the framework provides a comprehensive approach to assessing FAIRness, three limitations should be noted. Some aspects of the evaluation are inherently qualitative and may involve subjective judgment. This is due to the openness of the questions in the used questionnaire, and implies that the a repetition of the study might lead to slight differences in some aspects. Such nuances however would not affect the derived recommendations and future directions. Additionally, please note that the scope of the assessment is specific to the Digital Shadow Reference Model and may not take into account external factors affecting its FAIRness. As the FAIR Principles or the Digital Shadow Reference Model evolve, the relevance of some of the criteria may change, necessitating periodic updates to the contributions in this paper\footnotetext{\url{https://www.go-fair.org/how-to-go-fair/fair-implementation-profile/}}.

\section{Results and Analysis}

The results of our FAIRness evaluation of the Digital Shadow Reference Model are summarized in Table~\ref{table:findings}. 
The first table column refers to the questions from the FIP mini-questionnaire shown in Table~\ref{table:fair_criteria}.
The second column gives the answers to these questions from the perspective of the Digital Shadow Reference Model, and the third column indicates a rough measure derived from these answers.
Note that the last column does not represent a quantitative metric but rather represents a better illustration.

\begin{table*}[tbhp]
\caption{FAIRness assessment of the Digital Shadow Reference Model using the questions from Table~\ref{table:fair_criteria}.}
\begin{center}
\begin{tabular}{|l|l|c|}
\hline
\textbf{FIP Question} & \textbf{Digital Shadow Reference Model Answer} & \textbf{Evaluation (-{}-/-/o/+/++)} \\ \hline
\#1                                  & None                                      & -{}-                         \\ \hline
\#2                                  & HTTP URIs/URLs                                      & o                          \\ \hline
\#3                                  & This model is a metadata schema itself (no taxonomy or PIDs used)                                      & +                          \\ \hline
\#4                                  & Included custom properties such as \emph{data trace} or \emph{digital/physical asset}                                      & +                         \\ \hline
\#5                                  & None                                      & -{}-                          \\ \hline
\#6                                  & Project-internal tools like \emph{Dataset Finder}                                      & o                         \\ \hline
\#7                                  & HTTP REST or manual file transfer                                      & o                         \\ \hline
\#8                                  & Protocols such as HTTP / FTP / Git linked in a lightweight manner                                      & o                         \\ \hline
\#9                                  & Project-internal repositories                                      & o                          \\ \hline
\#10                                 & HTTPS / FTPS / Git with credentials                                      & +                          \\ \hline
\#11                                 & No explicit plan. Some assets are tracked with Git                                      & -                          \\ \hline
\#12                                 & Loosely structured UML or JSON                                      & -                          \\ \hline
\#13                                 & Free choice; no constraints or suggestions                                      & -                         \\ \hline
\#14                                 & None                                      & --                         \\ \hline
\#15                                 & DCAT derivative with uncontrolled entries & -                         \\ \hline
\#16                                 & UML and custom cross-references                                      & o                         \\ \hline
\#17                                 & Hierarchical references (project structure)                                      & -                        \\ \hline
\#18                                 & Not covered                                      & -{}-                         \\ \hline
\#19                                 & Not covered                                      & -{}-                         \\ \hline
\#20                                 & Custom provenance information                                     & o                         \\ \hline
\#21                                 & Custom fields based on data management plans                                      & o                         \\ \hline
\end{tabular}
\label{table:findings}
\end{center}
\end{table*}

The areas of strength are the answers to questions \#3, \#4 and \#10, which address the \emph{findability} and \emph{accessibility} areas of the FAIR Principles.
These were evaluated with a \texttt{(+)} score, because the Digital Shadow Reference Model itself is a metadata schema, it includes custom properties such as \emph{data traces} or \emph{digital/physical assets} for linking between data and metadata, and it supports authentication and authorization techniques such as HTTPS, FTPS or Git credentials.

Neutral \texttt{(o)} scores were assigned to multiple answers, because the Digital Shadow Reference Model's degree of alignment with the FAIR Principles are limited.
These include the support of basic HTTP URIs as dataset identifiers (\#2) and a project-internal tool for indexing datasets (\#6) from the \emph{findable} area.
From the \emph{accessible} area, multiple answers had a neutral score: The limited support for standardized communication protocols for data and metadata (\#7 and \#8), and the support for authentication and authorization techniques through project-internal repositories only.
In the \emph{interoperability} area, a neutral score was achieved because it supports schemas for metadata records (\#16) but only via UML and custom cross-references.
In the \emph{reusability} area, two neutral scores were achieved for supporting provenance information for both data and metadata (\#20 and \#21), but only with custom fields.

The main weaknesses are as follows.
No globally unique, persistent, resolvable identifiers are used for metadata records (\#1), and metadata records are not indexed (\#5) in search engines.
Longevity is only ensured by tracking some assets with Git, but no explicit longevity plan is present (\#11).
With regards to \emph{interoperability}, neither machine-readable knowledge representation languages nor structured vocabularies are used for data or metadata (\#12 to \#15).
The only available dataset schema (\#17) is a hierarchical structure based on the project structure.
A proper usage of licenses for data or metadata assets is not covered (\#18 and \#19).

In summary, the results of the evaluation show varying degrees of alignment between the Digital Shadow Reference Model and the FAIR Principles. Strengths were identified in the areas of findability and accessibility, with positive scores reflecting the model's use of custom properties for linking data and metadata, and its support for authentication and authorization mechanisms. However, the assessment also identified several aspects with neutral scores, including basic record identifiers, limited indexing tools, and limited support for standardized protocols, schemas, and provenance information. These results suggest that while the model includes some FAIR-aligned features, many are only partially realized or implemented in a project-specific manner.

A more detailed analysis reveals notable gaps that hinder the model's broader FAIRness. Key weaknesses include the lack of globally unique, resolvable identifiers, inadequate indexing of metadata, and insufficient planning for asset longevity. Interoperability is constrained by the lack of machine-readable languages, structured vocabularies, and robust schemas for datasets, while data reusability is limited by the lack of explicit licensing and alignment with domain standards. These gaps underscore critical areas where the model could evolve to achieve a better FAIRness.

\section{Recommendations and Future Directions}

To improve the FAIRness of the Digital Shadow Reference Model, several targeted improvements are recommended. First, the adoption of globally unique and resolvable identifiers, such as Digital Object Identifiers (DOIs) or PIDs, would greatly improve the findability of data and metadata. The implementation of standardized metadata schemas, using Web standards such as RDF or JSON-LD, could ensure broader interoperability and better alignment with the FAIR Principles. Indexing metadata in (public) search engines and registries would facilitate greater accessibility. In addition, the adoption of standardized communication protocols and structured vocabularies or ontologies, along with formalized licensing of data and metadata assets, would enhance both accessibility and reusability. Together, these measures aim to bridge existing gaps and support making the model more FAIR-compliant.

One step forward would be to release an updated version of the Digital Shadow Reference Model that explicitly addresses the identified gaps. This new version should prioritize features such as globally unique identifiers, formalized vocabularies, and metadata schemas that conform to Web standards such as RDF or JSON-LD. Collaborative workshops or focus groups could engage stakeholders to validate these updates and ensure alignment with industry and academic needs. These steps must continuously consider practical feasibility of any proposed improvement. To measure progress, periodic re-evaluations using the FAIR evaluation framework are essential to provide transparency and drive iterative improvements. Future efforts could also include the development of implementation guidelines for practitioners and case studies that demonstrate the impact of the improved model on real-world IoT, IIoT, and IoP applications. These steps would not only improve FAIR compliance, but also position the model as a benchmark for similar frameworks.

\section{Conclusion}

This paper assessed the FAIRness of the Digital Shadow Reference Model, a key framework for managing and integrating data in IoT, IIoT, and IoP systems. Using a structured assessment based on the FAIR Data Principles and FAIR Implementation Profiles, we identified areas of strength, such as metadata linking and support for authentication mechanisms, as well as significant gaps, including the lack of globally unique identifiers, limited use of standardized protocols, and insufficient metadata indexing.  

To address these gaps, we proposed recommendations, including the adoption of machine-readable metadata schemas, standardized vocabularies, public metadata indexing, and formal licensing practices. By implementing these improvements and iteratively evolving the model, the Digital Shadow Reference Model can better align with the FAIR Principles, increasing its utility and impact across diverse digital ecosystems. This work lays a foundation for future efforts to integrate the FAIR Principles into reference models to promote more efficient, collaborative, and sustainable data practices.

\section*{Acknowledgment}

Funded by the Deutsche Forschungsgemeinschaft (DFG, German Research Foundation) under Germany's Excellence Strategy – EXC-2023 Internet of Production – 390621612.
During the preparation of this work, we used DeepL and OpenAI's GPT-4o to improve the writing, make suggestions, and for rephrasing. After using these solutions, we carefully reviewed and edited the content as needed. The author takes full responsibility for the content of this article.

\bibliographystyle{IEEEtran}
\bibliography{references}

\end{document}